\documentstyle[aps,prl,epsf,floats]{revtex}
\begin{document}
\title{The relaxation of initial condition in systems with infinitely many
absorbing states}
\author{G\'eza \'Odor$^1$, J.F. Mendes$^2$, M.A. Santos$^2$ and M.C. Marques$^2$\\}
\address{$^1$Research Institute for Technical Physics and Materials Science,
          P. O. Box 49, H-1525 Budapest, Hungary}
\address{$^2$ Departamento de F\'\i sica and Centro de F\'\i sica do Porto, 
Faculdade de Ci\^encias, Universidade do Porto\\
Rua do Campo Alegre, 687 - 4150 Porto -- Portugal}
\date{\today}
\maketitle
 
\begin{abstract}
We have investigated the effect of the initial condition on the spreading
exponents of the one-dimensional pair contact process (PCP) and 
threshold transfer process (TTP).The non-order field was found to exhibit critical 
fluctuations, relaxing to its natural value with the same power-law as the order 
parameter field. We argue that this slow relaxation, which was not taken into 
account in earlier studies of these models, is responsible for 
the continuously changing survival probability exponent.
High precision numerical simulations show evidence of a 
(slight) dependence of the location of the transition point on the initial 
concentration, in the case of PCP.
The damage spreading (DS) point and the spreading exponents coincide with 
those of the ordinary critical point in both cases. 
\end{abstract}

\section{Introduction}

Recently the question was addressed  whether one can construct
initial states that affect the {\it entire} temporal evolution of critical
non-equilibrium systems. 
This is the case of systems that display a phase transition between an 
active state and a phase with infinitely many absorbing states. 
Non-universality of dynamic properties, associated with the initial 
configuration dependence of the survival probability of 
clusters started from a single active site, has been reported 
\cite{Jensen-Dickman,Mendeshyp,Dick2D}.

Similar changes in the critical spreading behaviour have been observed by 
Grassberger et al \cite{GCR} in a model where long-time memory effects are explicitly
introduced. In the system studied by these authors, the susceptibility to the 
spreading of an active agent changes after the first encounter, remaining constant 
afterwards. Despite the observed non-universal dynamical critical behaviour, 
susceptibility in the first encounter does not affect the critical point location. 
Grassberger et al argue that these results apply to models with multiple absorbing 
states where an effective memory-dependent susceptibility is present.
However, different behaviour is predicted in the case of slowly decaying memory 
effects, in which case both the value of the critical point and the exponents 
are expected to depend on the initial state.

A dependence on the initial configuration has also been found 
\cite{hayecikk} in the case of long-range spatial correlations, in which 
case the dynamical critical exponents change continuously as a function 
of initial correlation length.

The critical relaxation from an initial homogeneous state in systems 
with multiple absorbing states was recently examined from a field theoretical
(Langevin equations) approach \cite{MGDL}. The evolution equation for 
the order parameter density was found to include a memory term which 
is not present in the simpler case of directed percolation (DP).

In the present study we have investigated memory effects in two 1-d models with 
multiple absorbing states, the PCP \cite{Jensen-PCP,Jensen-Dickman} and 
TTP \cite{Mendeshyp} models. In both models there is a non-order parameter
field, dynamically coupled to the order parameter field, that gives rise to 
an effective susceptibility for spreading. We show that the non-ordering field 
relaxes to its steady state value by the same power-law time dependence as 
the order parameter field and this is characterised by the natural, long-time 
behaviour exponent of the density decay of a DP process. This is clear 
evidence of slowly decaying memory.
This power-law boundary condition (in time) is similar to that of the
long-range power-law boundary condition (in space) of \cite{hayecikk} and we
can see the emergence of continuously changing dynamical critical exponents.
The small shift of the critical point as a function of initial conditions, shown 
by high precision simulations for the PCP model, is in agreement with the above 
arguments.

Damage spreading (DS) simulations invented in biology \cite{Kauf} and later in
physics \cite{damphys} are useful to show the stability of the systems with
respect to small perturbations. The spreading behaviour has been shown to be
sensitive to the dynamics leading to the same steady states. An "objective"
definition of DS has been proposed \cite{Dom-Haye} according to which the
phase diagrams of the steady states of non-equilibrium models can be divided
to sectors in which all, none or parts of the physically possible dynamical 
rules generate stable damages. The phase transitions between the phases can be
continuous and usually belong to the DP universality class \cite{GrasDS}.
However, if the damage variables possess conservation and the DS absorbing 
states exhibit symmetries - which usually happens when the DS transition 
point coincides with the ordinary critical point the DS transition can 
belong to a different universality class \cite{Haye-DS,Od-ME-DS}. 
The DS transition cannot be in the passive phase of the replicas, but 
if it occurs in the active phase, the fluctuating replicas at the DS 
absorption point exclude the non-DP DS behaviour \cite{Od-ME-DS}. 
We have investigated the DS properties of the PCP and TTP models; 
the DS transition is shown to coincide with the ordinary critical point, 
and the non-universal spreading exponents have been inherited as well.

In section $II$, we give a brief introduction to the PCP and TTP models.
Time-dependent and critical relaxation studies are described in $III$, 
whereas section $IV$ is devoted to damage spreading simulations. 
Some comments and conclusions are presented in $V$. 

\section{The PCP and TTP models}

Both of these models have a single control parameter $p$ and qualitatively similar 
phase diagrams, displaying an 'active state' ($p<p_c$ in case of PCP, $p>p_c$ for TTP)
and  (infinitely many) absorbing phases ($p>p_c$ in case of PCP, $p<p_c$ for TTP).

In the TTP model, each site may be vacant, single or doubly (active) occupied,
and this can be described by a $3$-state variable $\sigma_{i} =0,1,2$. In
each time step, a site is chosen at random. In the absence of active sites, 
the dynamics is indeed trivial: if $\sigma_{i}(t)=0$ (or $1$), then
$\sigma_{i}(t+1)=1 (0)$ with probability $p$ (or $1-p$). The system relaxes 
exponentially to a steady state where a fraction $p$ of sites have $\sigma_{i}=1$ 
and the others are vacant. 
If $\sigma_{i}(t)=2$, then $\sigma_{i}(t+1)=0, \sigma_{i-1}(t+1)=\sigma_{i-1}(t)+1,
\sigma_{i+1}(t+1)=\sigma_{i+1}(t)+1$ if $\sigma_{i+1}(t)$ and $\sigma_{i-1}(t)$ are 
both $<2$ and $\sigma_{i}(t+1)=1$ if only one
of the nearest neighbours of site $i$ ($j=i-1$ or $i+1$) has $\sigma_{j}(t)< 2$,
in which case $\sigma_{j}(t+1) =\sigma_{j}(t)+1$.
As can be easily seen, the number of active sites either decreases or
remains the same in all processes other then $(1,2,1) \rightarrow (2,0,2)$;
 the
frequency of these processes depends on the concentration of '1'-s, which is 
controlled by the parameter $p$. Any configuration consisting of only '0'-s or 
'1'-s is absorbing in what concerns the active sites. The absorbing states in this 
model are fluctuating - in the respective sector of phase space, ergodicity is not
broken.

As we show below, the dynamics of '1'-s is however strongly affected by the presence 
of active sites. At the critical point, the concentration of '1'-s relaxes
to its steady state value (equal to $p_ {c}$) by a power-law.

The PCP is a $2$-state variable model with multiple absorbing states, each one
of them completely frozen in time, contrary to what happens in the TTP.
In the PCP, nearest-neighbour pairs of particles (dimers or active sites) annihilate 
each other with probability $p$ or create, with probability $1-p$, a particle at
one of the adjacent (vacant) sites to the dimer. Dimers cannot be generated 
spontaneously and therefore play the role of the '2'-s in the TTP. There is a natural
configuration to which the system at
criticality evolves after all activity has died out; the relationship between
 the natural particle density and $p_c$ is not a simple one unlike the TTP case.  
As shown below, the relaxation
to this natural state is a slow process - a power-law in time is also found
in this case and slowly-decaying memory effects arise as a result of
the coupling between the local density of dimers and the local density
of isolated occupied sites. 

\section{Time Dependent Simulation results}

Time dependent simulations \cite{GrasTor} have become an effective tool 
to explore dynamical critical exponents of systems at non-equilibrium 
phase transition points. The simulations are started from a single active
seed embedded in a sea of inactive sites and followed up to some $t_{MAX}$
time such that the cluster size can not exceed the system size $L$.
The quantities usually investigated are the mean number of active
sites $N(t)$ (pairs in case of PCP and '2'-s in case of TTP model) 
averaged over all trial samples, the survival probability $P(t)$ of the 
clusters and the mean spreading size $R(t)$ of the surviving clusters.
At the critical point and for asymptotically long times these quantities
exhibit power-law behaviour like 
\begin{equation}
N(t)\propto t^{\eta}
\end{equation}
\begin{equation}
P(t) \propto t^{-\delta}
\end{equation}
\begin{equation}
R(t)\propto t^{z/2} \ ,
\end{equation}
which define the exponents $\eta$, $\delta$ and $z$ respectively. The cluster 
size exponent $z$ characterising the linear scale is related to the anisotropy 
exponent of the system $ Z = \nu_{||}/\nu_{\perp}$  by $Z = 2 / z$. 
The order-parameter density inside the surviving clusters can be expressed in 
terms of these exponents as
\begin{equation}
\rho (t) \propto t^{\eta + \delta - d z/2}
\end{equation}
and is expected to show the same long-time decay as from an arbitrary bulk 
configuration with $\rho(0) \neq 0$, i.e.
\begin{equation}
\rho (t) \propto t^{-\beta / \nu_{||}}
\end{equation}
where $\beta$ is the steady state order-parameter exponent.

Exponents $\eta$ and $\delta$ are found to depend on the initial concentration 
of particles $\rho_1(0)$, and are related to static exponents $\beta$ and 
$\nu_{||}$ by the hyperscaling relation \cite{Mendeshyp}

\begin{equation}
2\eta + 2(\delta + \beta / \nu_{||}) = d z. 
\end{equation}

Seed growing simulations for the PCP and TTP models have been carried out up to
$t_{MAX} = 8000-16000$ time steps for $2\times 10^5$ trial runs.
We measured the order parameter density $\rho_2 (t)$ as well as the relaxation of 
the non-order field density $\rho_1(t)$ 
towards the natural values $\rho_1^{nat} \equiv \rho_1(\infty)$ of the models.
In case of the PCP we used $\rho_1^{nat} = 0.242(1)$ \cite{DickSOC}
while for the TTP model  $\rho_1^{nat} = p_c = 0.6894(3)$ \cite{Mendeshyp}.
The densities were measured inside the "infected" regions of surviving
clusters only.
To estimate the critical exponents and the transition points together,
we determined the local slopes of the scaling variables.
For example, in the case of the order parameter density we computed

\begin{equation}
-\alpha(t) \equiv {\log \left[ \rho_2(t) / \rho_2(t/m) \right] \over \log(m)}
\end{equation}
with $m=8$. When $p = p_c$, one should see a horizontal straight line as
$1/t \to 0$. The off-critical curves should possess curvature:
curves corresponding to $p > p_c$ should veer upward, curves with $p < p_c$
should veer downward.

\begin{figure}[h]
\epsfxsize=100mm
\centerline{\epsffile{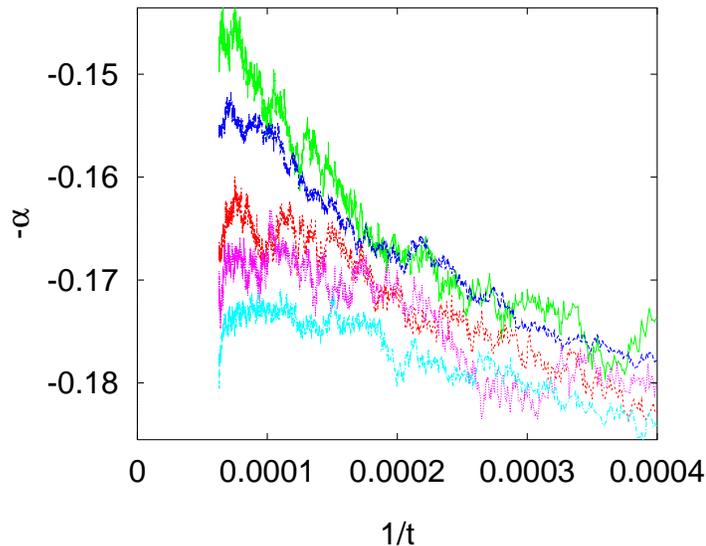}}
\vspace*{0.5cm}
\caption{Local slopes of $\Delta\rho_1(t)$ in PCP simulations at 
$\rho_1(0)=0$ for $p=0.077, 0.07695, 0.0769, 0.07683, 0.07675$ 
(from bottom to top curves). 
Scaling can be observed for $p=0.07685(5)$, with the exponent 
$\alpha(t\to\infty)\approx 0.16$.}
\vspace*{4mm}
\label{PCP-TDS-1}
\end{figure}

\begin{figure}[h]
\epsfxsize=100mm
\centerline{\epsffile{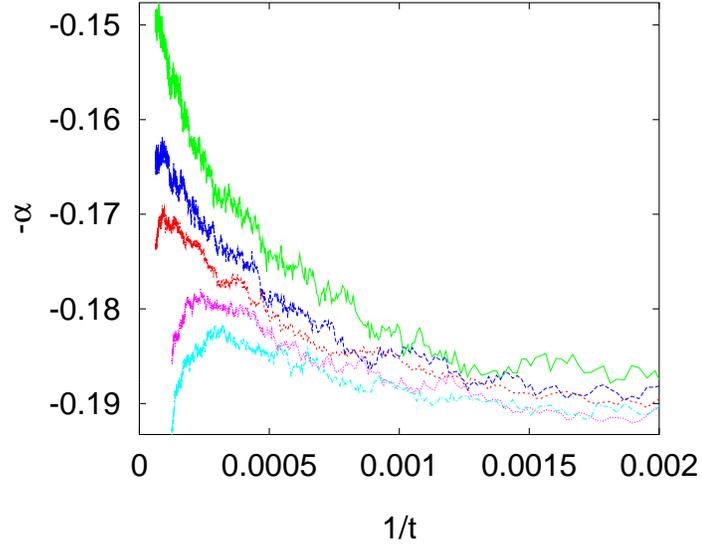}} 
\vspace*{0.5cm}
\caption{The same as Figure \ref{PCP-TDS-1} in case of $\rho_1(0) = 0.432$
and $p=0.07715, 0.0771, 0.07705, 0.077, 0.0769$
(from bottom to top curves). Scaling can be observed for $p=0.0770(5)$, 
with the exponent $\alpha(t\to\infty)\approx 0.16$.}
\vspace*{4 mm}
\label{PCP-TDS-2}
\end{figure}

Figures \ref{PCP-TDS-1} and \ref{PCP-TDS-2} show the local slopes of 
$\Delta \rho_1(t) \equiv  \rho_1(t) - \rho_1^{nat}$ for the PCP in case of 
$\rho_1(0)=0$ and $\rho_1(0)=0.432$ respectively. For the order parameter density 
we obtained the same results within numerical accuracy.
As one can read off, the particle density exhibits long-time power-law 
behaviour with DP exponent, but the critical point is slightly lower in case 
of $\rho_1(0)=0$ than in case of $\rho_1(0)=0.432$.
This however agrees with the slow-relaxing susceptibility picture, 
because $\rho_1(0)=0$ is smaller than the natural value $\rho_1^{nat}=0.242(1)$
and increases very slowly in the bulk, therefore a smaller annihilation
probability is enough to drive the system to the absorbing state.
This dependence can also be the result of the lattice version of the
PCP model, where, in contrast to the field theory where there is always some
finite density of fields, the isolated '1'-s are frozen.
Both densities exhibit the same exponents in  good agreement with 
$\beta / \nu_{||} = 0.1596(4)$ of the $1+1$ d DP class.
The survival probability exponent measured in our high precision simulation is 
in agreement with the value of \cite{Jensen-Dickman} within numerical accuracy.

In the critical point estimates we could see a tiny difference depending on 
whether we measured the survival probability or the density in the same simulation.
Namely the $p_c$ appears to be nearer to the results of \cite{Jensen-Dickman,Dpriv} 
if we estimated it from the $P(t)$ data ($p_c=0.07704$ instead of $p_c=0.07685$ 
in case of $\rho_1(0)=0$ and $p_c=0.07714$ instead of $p_c=0.0770$ in case of
$\rho_1(0)=0.432$).
This small offset can be understood on the basis of our density measuring method
plus the slow, frozen relaxation in PCP that seems to cause a crossover effect. 
The density has been averaged in the infected regions, where frozen
"islands" can appear which don't evolve at all but keep the non-natural
densities for long times. Therefore we overestimate the size of the region in 
which densities really relax. To verify this picture we have performed simulations
where we averaged the '1's and '2' over fixed size. In this case we started the
PCP process from randomly distributed pairs in a system of size $L=16000$ and
followed the evolution up to $t_{MAX}=16000$. In cases of $\rho_1(0)=0$ and
$\rho_1(0)=0.33$ we found that the critical point is about $p_c=0.07708$ nearer to the
$P(t)$ and the $N(t)$ DS results. The scaling exponent was again DP like : 
$\alpha\approx 0.16$.

\begin{figure}[h]
\epsfxsize=100mm
\centerline{\epsffile{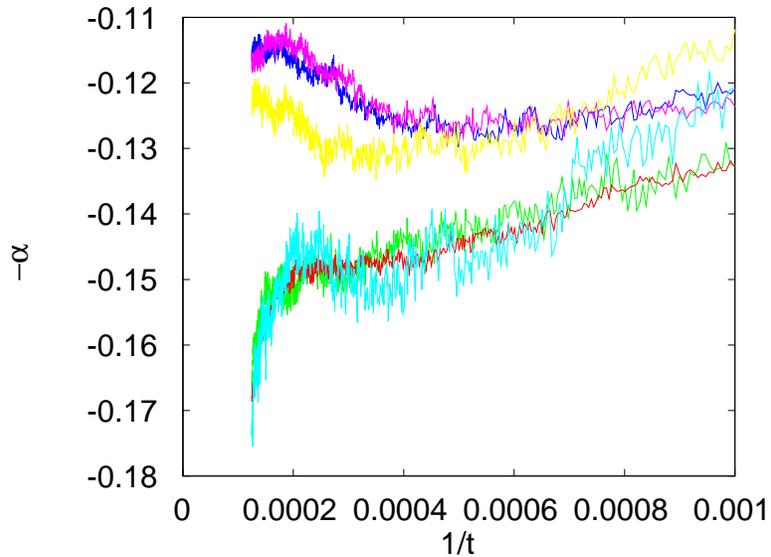}}
\caption{Local slopes for $\Delta\rho_1(t)$ in TTP simulations
for $p=0.6887$ (top curves), $p=0.6897$ (bottom curves);
and $\rho_1(0) = 0.4, 0.6894, 0.8$.
Scaling can be observed for about $p=0.6894$, with the exponent 
$\alpha(t\to\infty)\approx 0.16(1)$.}
\vspace*{4mm}
\label{TTP-TDS-1}
\end{figure}

In the case of TTP model we have performed simulations for $\rho_1(0)=0.4$,  
$\rho_1(0)=0.6894$ and $\rho_1(0)=0.8$ initial particle densities. 
In this model, we did not observe any shift of $p_c$, as Figure 
\ref{TTP-TDS-1} shows, and all the densities scale with the 
$\beta / \nu_{||} = 0.1596(4)$ exponent.

The similar power-law behaviours of $\rho_2 (t)$ and $\Delta\rho_1(t)$ 
can be understood if one considers the coupled Langevin equations describing 
the time evolution of the processes.
For example, in case of the PCP model, they look like \cite{MGDL}:

\begin{equation}
{\partial \phi_1(x,t) \over \partial t} = c_1 \nabla^2_x \phi_2 + r_1\phi_2 
- u_1\phi_2^2 - w_1\phi_1\phi_2 + ... + \eta_1(x,t) \label{eq8}
\end{equation}
\begin{equation}
{\partial \phi_2(x,t) \over \partial t} = c_2 \nabla^2_x \phi_2 + r_2\phi_2 
- u_2\phi_2^2 - w_2\phi_1\phi_2 + ... + \eta_2(x,t)  \label{eq9}
\end{equation}

where $\eta_1(x,t)$ and $\eta_2(x,t)$ are Gaussian uncorrelated noise terms 
proportional to $\sqrt {\phi_2}$.
One can see that the equations are coupled strongly by $w$-terms; since the right hand 
sides of the equations contain the same powers of the scaling fields, the time 
derivatives are expected to have the same scaling too.

In reference \cite{MGDL} it was argued that

\begin{equation}
\phi_1(t) = \phi_1^{nat} +(\phi_1(0)-\phi_1^{nat})e^{-w_1\int_0^t \phi_2(x,s) ds}
\label{eq10}
\end{equation}

($\phi_1^{nat} = r_1/w_1$ is the natural concentration in this formalism) may be 
taken as an approximate solution of equation (\ref{eq8}), in which case the 
$\phi_1\phi_2$ cross-term in (\ref{eq9}) has the form

\begin{equation}
-w_2 \phi_ 2 (\phi_1(0)-\phi_1^{nat})e^{-w_1\int_0^t \phi_2(x,s) ds}. \label{eq11}
\end{equation}

The power-law time dependence of $\phi_1$ 
is obvious, because an exponential relaxation to $\phi_1^{nat}$ would just
give in (\ref{eq9}) a term similar to $r_2\phi_2$ that can shift the critical 
point but not the critical indices (from the DP values).

\begin{figure}
\epsfxsize=100mm
\centerline{\epsffile{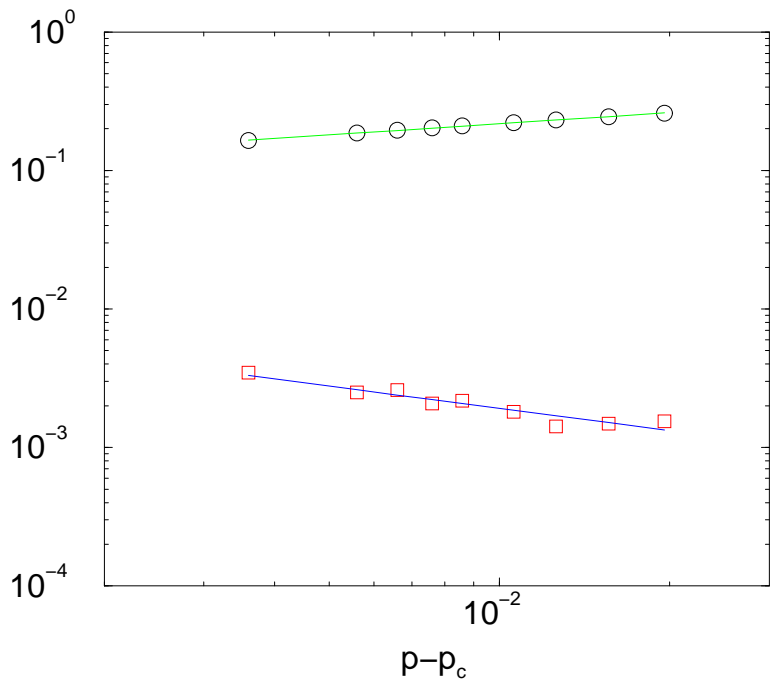}}
\caption{Log-log plot of $\rho_1(p) - \rho_1(p_c)$ ($\bigcirc$)
and respective fluctuation ($\Box$) v.s. $p-p_c$ above the critical 
point of the TTP model. 
Least-square regression results in $\beta = 0.27(2)$ and $\gamma=0.53(6)$. 
The averaging was done following 45000 initialisation MC steps over 5000 time 
steps and 300 trials.}
\label{steady}
\end{figure}

The long-range scaling behaviour of the non-order density suggests that the
$\phi_1$ field possesses critical fluctuations. To test this, we have performed
steady state simulations as well. We have measured the '0', '1' and '2' densities
in case of the TTP model just above the critical point. We considered $L=4000$ 
systems and let them evolve from random initial conditions with $p$ slightly 
above $p_c=0.6894$; about 40000 MC lattice updates 
were necessary to reach the steady state. As one can see in Figure \ref{steady}, 
least-square fits of $\log (\rho_1(p) - \rho_1(p_c) )$ v.s. $\log (p - p_c)$ 
resulted in regular $DP$ scaling exponent $\beta=0.27$ \cite{dimer}. The other 
two densities ('0'-s and '2'-s) exhibited the same steady state exponents too.
For the fluctuation of $\Delta\rho_1 \equiv \rho_1(p) - \rho_1(p_c)$ 
we observed the scaling 
\begin{equation}
\langle\Delta\rho_1^2\rangle - \langle\Delta\rho_1\rangle^2 \propto |p-p_c|^{-\gamma} 
\end{equation}
with $\gamma=0.53(6)$ exponent, which agrees with DP universality class value
again \cite{Jensen96}. 

\begin{figure}[h]
\epsfxsize=100mm
\centerline{\epsffile{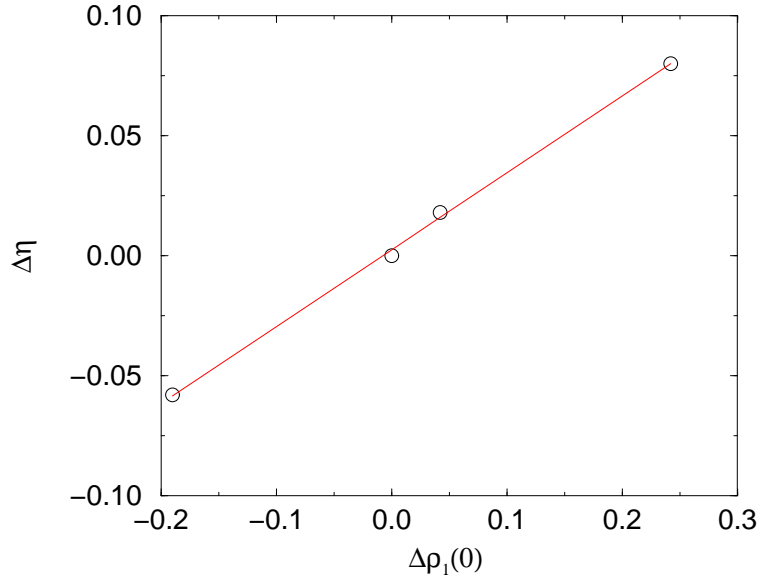}}
\caption{Initial concentration dependence of the exponent $\eta$ for
PCP model. Linear regression gives a slope $0.320(7)$ between $\eta-\eta_{DP}$
and $\rho_1(0)-\rho_1^{nat}$}
\label{pcp_res}
\end{figure}

The critical behaviour of the $\phi_1$ field demands that extra care is taken when 
dealing with truncated versions of equations (\ref{eq8}) and (\ref{eq9}).
A numerical integration of (\ref{eq9}), including the non-Markovian term 
(\ref{eq11}), was carried out by Lopez and Mu\~noz \cite{lopez} and 
revealed that the presence of the memory term is responsible for
scaling up to some time with non-universal values of $\eta$ 
and $\delta$. However, these authors did not find the linear relation
between the shift from the DP values ($\eta-\eta_{DP}$ and
$\delta-\delta_{DP}$) and $\phi_1(0)-\phi_1^{nat}$ that our results for PCP
show (Figure \ref{pcp_res}) and was also found in previous TTP studies 
\cite{mendesthesis}. We think one has to take into account the 
omitted terms in eq. (\ref{eq10}), which we have shown to exhibit power law 
in time and therefore give a relevant contribution to the 
renormalization of $\phi_2$.  
\begin{figure}[h]
\epsfxsize=100mm
\centerline{\epsffile{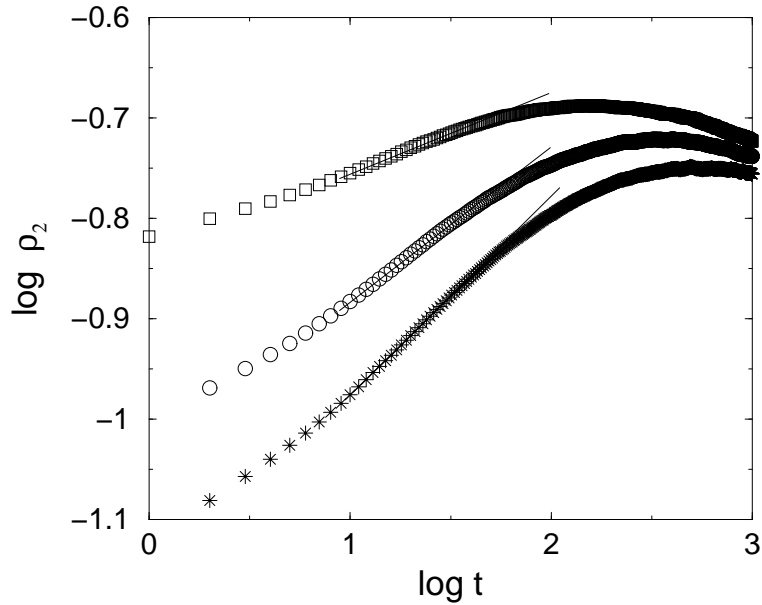}}
\caption{Double-logarithmic plot of $\rho_2$ v.s.$t$ for critical TTP 
in early times for $\rho_2(0)=0.02$ and $\rho_1(0)=0.4, 0.69$ and $0.8$ 
(top to bottom curves).The slopes of the straight lines are $0.08,0.15$ 
and $0.19$ (respectively).}
\label{slip}
\end{figure}

We have also investigated the early stages of the relaxation of 
the order parameter.\\
In reference \cite{MGDL} it was suggested that in the short time regime 
($w_{1} \phi _{2}t \ll 1$) one might observe the dynamic percolation 
scaling of Grassberger \cite{GCR}. Early time scaling - currently referred 
to as {\it critical initial slip} - was introduced by Jansen et al 
\cite{Jansen} and recently investigated by van Wijland et al \cite{Wijland}
for a reaction-diffusion model with two kinds of particles, using RG analysis. 
They found non-universal dependence (on the initial particle distribution) of the slip
exponent in the case of unequal diffusion coefficients.\\
The system was prepared by 'adding' a few '2'-s (uniform density $\rho_{2} (0) \ll 1$)
to a random uniform background of '0'-s and '1'-s (density $\rho_{1} (0)$). The system
evolution was recorded up to $t=1000MCS$ and averages were performed over 
independent runs ($10^{4}$ typically). Log-log plots of $\rho_{2}(t)$ - see 
Fig. \ref{slip} - show a linear region (lasting for $10  < t < 100$ for $\rho_{2}(0)$ 
equal to a few percent) with a slope that is independent of $\rho_{2}(0)$ but 
depends on $\rho_{1}(0)$.
This may be evidence for the critical initial slip with a non-universal slip exponent
$\theta^{'}$ equal to that slope. Unlike the DP case, where $\theta^{'}= \eta$ 
\cite{Wijland}, the values we found for $\theta^{'}$
differ significantly from the corresponding $\eta$ values.

\section{Damage spreading simulations}

\begin{figure}[ht]
\epsfxsize=100mm
\centerline{\epsffile{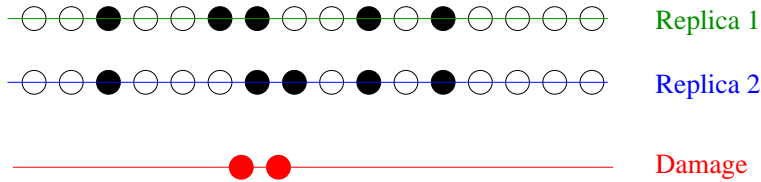}}
\vspace*{4mm}
\caption{Sample initial state of PCP DS simulations. 
Pairs (seeds) are displaced in the middle of the lattices of replicas with a 
single space shift that generates 2 damage variables. In case of TTP DS 
simulations the initialisation is the same, except that we have '2'-s instead of pairs.}
\label{DSini}
\end{figure}

\begin{figure}
\epsfxsize=100mm
\centerline{\epsffile{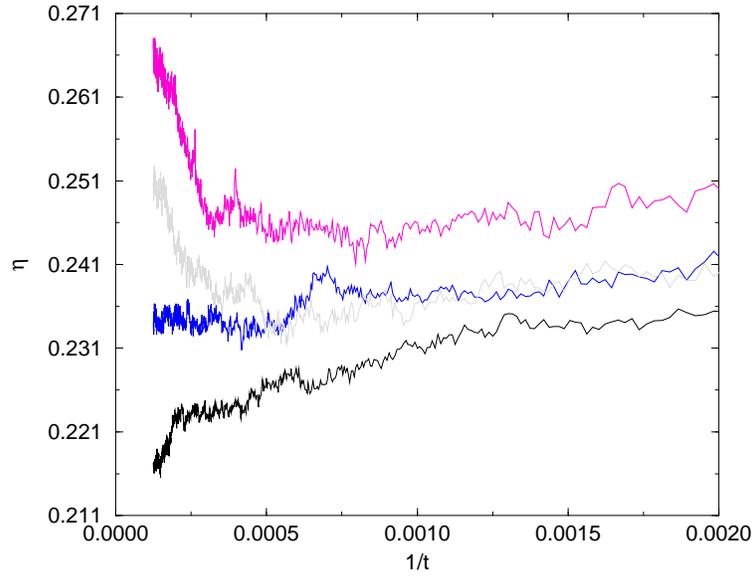}}
\caption{Local slopes $\eta(t)$ near the PCP DS transition point, for 
$\rho_1(0)=0$ and $p=0.07708, 0.07704, 0.077, 0.0769$ (from bottom to top). 
The DS critical point is at $p_d=0.07704$ with the corresponding exponent 
$\eta=0.234(3)$.}
\label{dam_0_eta}
\end{figure}

\begin{figure}
\epsfxsize=100mm
\centerline{\epsffile{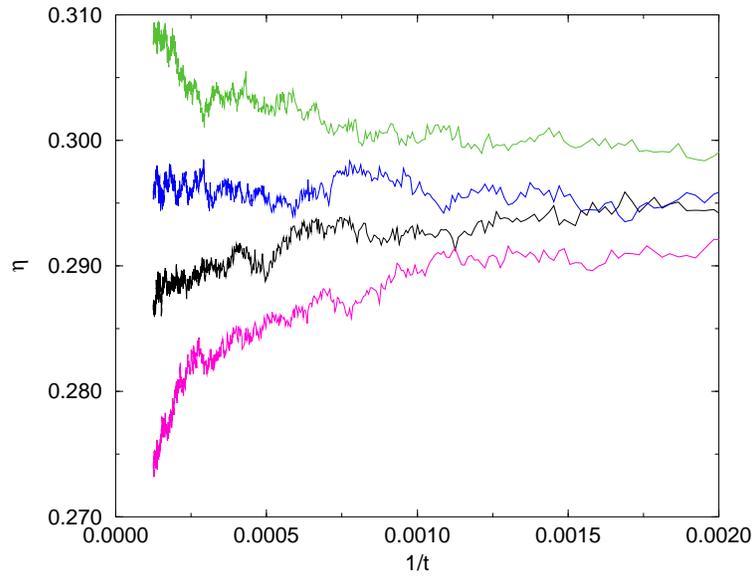}}
\caption{Same as figure \ref{dam_0_eta} for $\rho_1(0)=0.2$ and
$p=0.07716, 0.07712, 0.07708, 0.07704$ 
(bottom to top). The DS critical point is at $p_d=0.07708$ with 
the corresponding $\eta=0.296(1)$.}
\label{dam_2_eta}
\end{figure}

\begin{figure}
\epsfxsize=100mm
\centerline{\epsffile{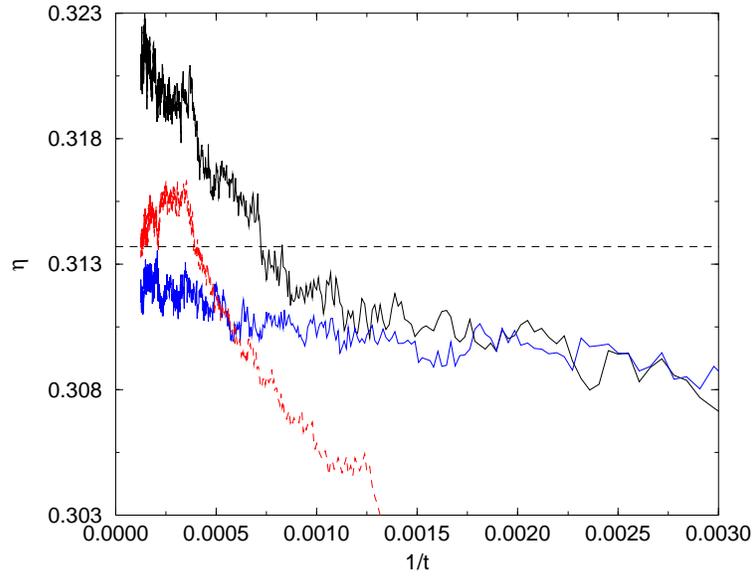}}
\caption{Same as figure \ref{dam_0_eta} for $\rho_1(0)=0.242$ and
$p=0.07709, 0.07707$ (from bottom to top). 
The dotted curve corresponds to simulations with system generated initial 
configurations and $p=0.07709$. 
The DS critical point is at $p_d=0.07709$ with the corresponding exponent
$\eta=0.314(6)$. The dashed line shows the estimated value of the DP exponent 
obtained by simulations.}
\label{dam_242_eta}
\end{figure}

\begin{figure}
\epsfxsize=100mm
\centerline{\epsffile{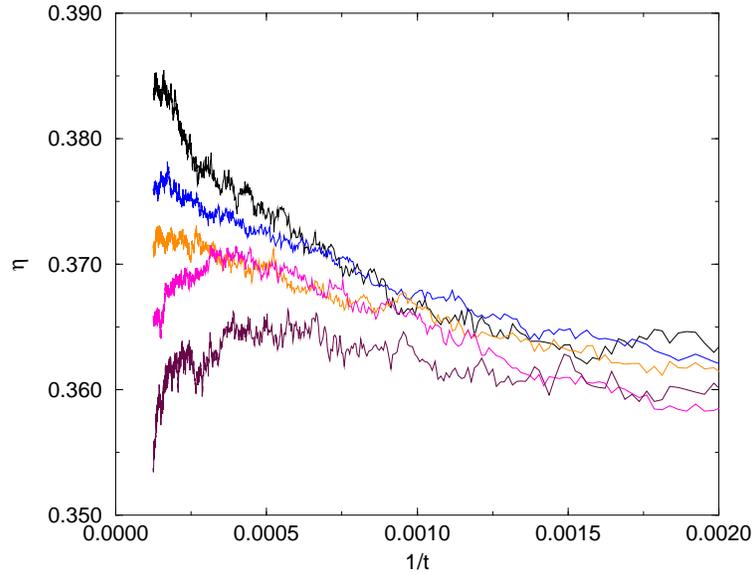}}
\caption{Same as figure \ref{dam_0_eta} for $\rho_1(0)=0.432$ and
$p=0.07718, 0.07716, 0.07714, 0.07712, 0.0771$ 
(from bottom to top). The DS critical point is at $p_d=0.07714$ with 
the corresponding exponent $\eta=0.372(5)$.}
\label{dam_432_eta}
\end{figure}

The damage spreading simulations have been initialised by two replicas of states
with identical, but random uniform distribution of single '1'-s of given 
concentrations. Then a seed (a pair in case of PCP and a '2' in case of TTP) is
added to each replica such that they become nearest neighbours and the initial 
difference is 2 (see fig.\ref{DSini}).
The order parameter characterising the damage is the Hamming distance between replicas
\begin{equation}
D(t) = \langle \sum_{i=1}^L \vert s(i) - s^,(i) \vert \rangle  \ .\label {Dscal}
\end{equation}
where $s(i)$ denote the pairs in case of PCP and the variable '2'-s in case of
the TTP model.
At the DS critical point ($p_d$) we expect that the order parameter scales as 
\begin{equation}
D(t)\propto t^{\eta} \ ,
\end{equation}
Similarly the survival probability of damage variables behaves as
\begin{equation}
P(t)\propto t^{-\delta} \,
\end{equation}
and the average mean square distance of damage spreading from the center scales
as
\begin{equation}
R^2(t)\propto t^z \ .
\end{equation}
Averages were performed over $N_s=10^6$ independent runs for each 
value of $p$ in the vicinity of $p_d$ ( but for $R^2(t)$ only over the surviving
runs). The $t_{MAX}$ was $8000$ in these simulations.
Figures \ref{dam_0_eta}, \ref{dam_2_eta}, \ref{dam_242_eta} and \ref{dam_432_eta}
show the local slopes results of the Hamming distance for the PCP model  
and different initial concentration of '1'-s. 
The $p_d$ transition points are found to coincide with the ordinary critical 
points of the replicas within numerical accuracy. A small, but monotonic 
tendency in the variation of $p_d$ (as in case of $p_c$) with initial conditions  
can be observed in Table I:

\begin{table}[ht]
\begin{tabular}{|c|c|c|c|c|}
$\rho_1(0)$     & $p_d$         & $\eta_{DS}$   & $\delta_{DS}$ & $z$ \\
\hline
$0.0$           & $0.07704$     & $0.234(3)$    & $0.24(2)$     & $1.23(3)$ \\
$0.2$           & $0.07708$     & $0.296(1)$    & $0.18(1)$     & $1.24(4)$ \\
$0.242$         & $0.07709$     & $0.314(6)$    & $0.16(9)$     & $1.24(5)$ \\
$0.432$         & $0.07714$     & $0.372(5)$    & $0.11(1)$     & $1.26(3)$ \\
\end{tabular}
\caption{\em Damage spreading simulation results in the PCP model}
\end{table}

The DS critical $\eta$ and $\delta$ exponents show the same non-universal 
behaviour as the corresponding ordinary critical exponents (see also 
\cite{DickSOC,Dpriv}) and coincide with them within numerical precision. 
The exponent $z$ is constant within numerical accuracy.

In the case of $\rho_1(0)=0.242$, we performed runs with uniform
initial distributions and with system generated configurations of 
isolated '1'-s with the same average concentration. The latter, 
non-uniform distribution was generated by letting a single replica to run 
at $p=0.07709$ until it reached the absorbing state; then the infected area 
of the system was used as initial state for DS simulations, similarly to what 
was done in ref. \cite{Jensen-Dickman}. We don't see significant differences 
(Fig. \ref{dam_242_eta}) between the two cases for $t$ large, both of them 
result in exponents in agreement with the best DP class $\eta_{DP}=0.3137$ 
value \cite{Jensen96}.
 
In case of the TTP model we performed DS simulations for $\rho_1(0)=0.4$ only.
We could see analogous DS behaviour as in the case of PCP. Again the critical
point and exponents coincide with the corresponding critical values.

Similarly to what has been observed in models that belong to the PC 
universality class \cite{Od-ME-DS},we can also conclude 
that if the DS transition point coincides with  $p_c$
the scaling behaviour is inherited.\\

An interesting implication of this result can be stated exploiting 
the possible mapping of these models to SOC models \cite{DickSOC}.
The corresponding critical sandpile models are not chaotic in the sense 
that the avalanches (or clusters) arising from the dropping of pairs 
(or seeds) to the lattice result in trajectories with 
power-law increasing differences only. This does not exclude the possibility
that other SOC models generated by the way of ref. \cite{DickSOC}
are chaotic, since if $p_d$ happens to be in the active phase, the perturbations 
in the SOC model generate differences that increase faster than a power law.

\section{Conclusions}

Two representatives of systems with a continuous phase transition to 
infinitely degenerate absorbing states (PCP and TTP models) have been investigated 
numerically in $1-d$. In order to clarify the influence of the initial 
condition on the dynamic properties, we have performed time-dependent 
simulations and analysed the evolution of both the order parameter and 
the non-order field densities. We gave numerical evidence that the 
non-order field is in a critical state simultaneously with the
order parameter field. The isolated particles density exhibits a continuous 
phase transition with DP exponents to a non-absorbing state, therefore
its fluctuations cannot be neglected when one tries to understand the 
non-universal behaviour of critical exponents.
Due to the dynamic coupling between the two fields, the slow (power-law)
decay of the background particle density induces a long-time 
memory of the susceptibility to spreading of the order parameter \cite{GCR}.

We have found that the estimates of $p_c$ obtained from the time
dependence of the survival probability are consistent with those given by
DS studies; a small shift as a function of $\rho_1(0)$ is exhibited, which we
interpret as due to the slowly decaying memory \cite{GCR}. The study of
the density in seed-growing simulations produced values of $p_c$ slightly
off the former ones; we think this is probably a crossover effect.

We don't see such a $p_c$ shift for the TTP model so this maybe
specific of the PCP model, connected to the non-ergodicity of its
absorbing states. 
In both models the critical exponents $\eta$ and $\delta$ have been 
found to behave linearly as a  function of the initial particle concentration.

Preliminary studies of the early time critical regime suggest the 
existence of a (short) initial slip regime characterised by an exponent 
which depends on the initial particle concentration  $\rho_1(0)$, 
in agreement with RG predictions for a similar model \cite{Wijland}. 
This topic needs however additional investigation.
  
The damage spreading investigations have shown that the DS point coincides
with the critical point and so the critical indices "inherit" the same
non-universal scaling behaviour similarly what was found in an
earlier study \cite{Od-ME-DS}.

We hope that our study will stimulate further field theoretical 
analysis of critical spreading in systems with many absorbing states.

\acknowledgements
We thank R. Dickman for very stimulating correspondence.\\
Support from the NATO grant CRG-970332 is acknowledged. This work was partially 
financed by Praxis XXI (Portugal) within project PRAXIS/2/2.1/Fis/299/94.
G. \'Odor gratefully acknowledges support from the Hungarian research fund OTKA 
(Nos. T025286 and T023552).\\
The simulations were performed partially on the FUJITSU AP-1000+
and AP-3000 parallel supercomputer.


\begin{references}
\bibitem{dimer} A similar result was obtained in  \cite{Jensen-Dickman} for 
the dimer-reaction model.
\end{references}
\end{document}